\documentclass[dvips,12pt,a4paper]{article}
\usepackage{epsfig}
\usepackage{a4wide}
\usepackage{amsmath}
\usepackage{latexsym}

\newcounter{figureno}
\newenvironment{capt}{
\phantom{mmmm}
\vspace*{6mm}
\parindent=0pt
\addtocounter{figureno}{1}
\begin{minipage}[t]{160mm}
\small\sl Figure~\thefigureno.\ }{\end{minipage}
\vspace*{-4mm}}
\begin{document}
\thispagestyle{empty}
\pagestyle{empty}
\renewcommand{\thefootnote}{\fnsymbol{footnote}}

\renewcommand{\thanks}[1]{\footnote{#1}}
\renewcommand{\thepage}{\arabic{page}}
\catcode`@=11
\def\citer{\@ifnextchar [{\@tempswatrue\@citexr}{\@tempswafalse\@citexr[]}}

%

\def\@citexr[#1]#2{\if@filesw\immediate\write\@auxout{\string\citation{#2}}\fi
  \def\@citea{}\@cite{\@for\@citeb:=#2\do
    {\@citea\def\@citea{--\penalty\@m}\@ifundefined
       {b@\@citeb}{{\bf ?}\@warning
       {Citation `\@citeb' on page \thepage \space undefined}}%
\hbox{\csname b@\@citeb\endcsname}}}{#1}}
\catcode`@=12
\newcommand{\nn}{\nonumber}

\newcommand{\dd}{\mbox{{\rm d}}}
\newcommand{\mH}{m_{\rm H}}
\newcommand{\dLips}{\mbox{{\rm dLips}}}

\def\Im{\mbox{\rm Im\ }}
\def\Re{\mbox{\rm Re\ }}
\def\fourth{\textstyle\frac{1}{4}}
\def\gsim{\mathrel{\rlap{\raise 1.5pt \hbox{$>$}}\lower 3.5pt
\hbox{$\sim$}}}
\def\lsim{\mathrel{\rlap{\raise 1.5pt \hbox{$<$}}\lower 3.5pt
\hbox{$\sim$}}}
\def\GeV{{\rm GeV}}

\def\KL{K_L}
\def\KS{K_S}
\def\cp{$CP$\ }
%
\def\Month{\ifcase\month\or
January\or February\or March\or April\or May\or June\or
July\or August\or September\or October\or November\or December\fi}
\def\slash#1{#1 \hskip -0.5em /}
%

\begin{flushright}
{\tt
\hfill
University of Bergen, Department of Physics \\
Scientific/Technical Report No.\ 1995-17 \\ ISSN~0803-2696\\
hep-ph/95mmnnn \\
November, 1995
}
\end{flushright}
\vspace*{1cm}

\vfill
\begin{center}
{\LARGE
$CP$ Investigations in the Higgs Sector\thanks{Invited paper,
{\em Third Tallinn Symposium on Neutrino Physics},
Lohusalu, Estonia, October 7--11, 1995}}

\vspace{4mm}
{\Large
Arild Skjold and Per Osland \\\hfil\\
\vspace{3mm}
Department of Physics\thanks{Electronic mail addresses:
{\tt \{skjold,osland\}@fi.uib.no}}\\
University of Bergen \\ All\'egt.~55, N-5007 Bergen, Norway
}
\end{center}
\date{}

\vspace{10mm}
In a more general electroweak theory, there could be
Higgs particles that are odd under $CP$.
When such particles decay via vector bosons to two
fermion-antifermion pairs, the momenta of those will be correlated
in a way which is determined by the \cp of the particle.
Similarly, in the Bjorken process correlations among momenta
of the initial electron and final-state
fermions are sensitive to the \cp quantum number.
Monte Carlo data on the expected efficiency demonstrate that it
should be possible to verify the scalar character of an intermediate-mass
Standard Model Higgs boson
after three years of data taking at a future linear collider.
This is most likely not possible at LEP2.
Signals of possible presence of $CP$
violation in the Higgs sector are briefly discussed.
\vfill
\pagestyle{plain}
\section{Introduction}
The only known \cp violation is in the $K$-meson system,
where so far five \cp violating observables have been studied.
These are the three ``classical" ones
\cite{CP}
\begin{eqnarray}
\eta_{+-}
&=&\frac{A(\KL\to\pi^+\pi^-)}{A(\KS\to\pi^+\pi^-)} \nn \\
\eta_{00}
&=&\frac{A(\KL\to\pi^0\pi^0)}{A(\KS\to\pi^0\pi^0)} \nn \\
\delta_l
&=&
\frac{\Gamma(\KL\to\pi^-l^+\nu_l)-\Gamma(\KL\to\pi^+l^-\bar\nu_l)}
     {\Gamma(\KL\to\pi^-l^+\nu_l)+\Gamma(\KL\to\pi^+l^-\bar\nu_l)}
\end{eqnarray}
and two more recently observed ones
\citer{Ramberg,Adler}
\begin{eqnarray}
\eta_{+-\gamma}
&=&\frac{A(\KL\to\pi^+\pi^-\gamma,\mbox{ \cp violating})}
        {A(\KS\to\pi^+\pi^-\gamma)} \nn \\
A_{00}(\tau)
&=&\frac{R(\bar{K^0}\to\pi^0\pi^0)(\tau)-R(K^0\to\pi^0\pi^0)(\tau)}
        {R(\bar{K^0}\to\pi^0\pi^0)(\tau)+R(K^0\to\pi^0\pi^0)(\tau)}
\end{eqnarray}
One hopes to also study \cp violation in the $B$ meson system,
at the planned SLAC $B$ factory and dedicated future detectors
at other facilities like HERA and the LHC.
The above observables (which are very small) are often expressed
in terms of $\epsilon$ and $\epsilon'$, e.g.,
\begin{eqnarray}
\eta_{+-}&\simeq&\epsilon+\epsilon' \nn \\
\eta_{00}&\simeq&\epsilon-2\epsilon' \nn \\
\delta_l&\simeq&2\,\Re\,\epsilon
\end{eqnarray}
with $|\epsilon|=(2.27\pm0.02)\cdot10^{-3}$
and $|\epsilon'|\ll\epsilon$.

It has been known since 1967 \cite{Sakharov}
that \cp violation is a necessary
prerequisite for creating a matter-antimatter asymmetry,
as the Universe appears to have.
But while the CKM matrix of the standard model naturally accomodates
\cp violation through the complex phase factors in charged-current
couplings of the quarks \cite{KoMa},
it appears that \cp violation through the CKM matrix
may not be sufficient to explain the baryon asymmetry of the Universe
\cite{Cohen},
so it is important to explore alternative sources.
Many years ago Weinberg suggested that \cp might be violated
in the Higgs sector, and presented an explicit model
where this is the case \cite{Wein76}.

We shall here explore the possibility of \cp violation in
the Higgs sector, following a purely phenomenological approach.
Of course, we assume a non-minimal Higgs sector.
Assuming one finds some Higgs-particle candidate, there are
two related questions of fundamental interest:
$(i)$~Is the Higgs candidate even or odd under $CP$? and
$(ii)$~Is \cp violated in the Higgs sector?

Two physical situations allow us to address these questions
in a phenomenological way:
\begin{align}
\label{eq:LHC}
&H \rightarrow V_1V_2\rightarrow (f_1\bar f_2)(f_3\bar f_4)
& &\mbox{(LHC)} \\
&e^+e^- \rightarrow ZH\rightarrow (f_1\bar f_1)H
& &\mbox{(NLC)}
\end{align}
In several ways, they are quite similar.

We adopt the following effective Lagrangian describing the $HVV$-vertex:
\begin{equation}
\label{eq:vertex}
{\cal{L}}_{HVV}
= 2\cdot 2^{1/4} \sqrt{G_{\rm F}}
\bigl[
m^{2}_V \ V^\mu V_\mu \ H
+ \fourth\, \eta \ \epsilon^{\mu \nu \rho \sigma}\
V_{\mu \nu} V_{\rho \sigma} \ H\bigr]
\end{equation}
where $V_{\mu \nu}=\partial_\mu V_\nu -\partial_\nu V_\mu$.
The first term describes the standard-model $HVV$ coupling
(even in $CP$), whereas the second term describes the coupling of
a $CP$-odd scalar to two vector bosons.
(In the MSSM \cite{MSSM}, this latter coupling is absent at tree level.)
Armed with this Lagrangian, we
address the above two questions in the following two sections.

\section{$CP$ Eigenstates}

Let us first consider Higgs decay, as it could presumably be studied
at the LHC:
\begin{equation}
H \rightarrow V_1V_2\rightarrow
[f_1(q_1)\bar f_2(q_2)][f_3(q_3)\bar f_4(q_4)]
\end{equation}
where $V_1V_2$ denote $W^+W^-$ or $ZZ$.
The differential decay rate can be written as
\citer{others,HagiwaraStong}
\begin{eqnarray}
\label{eq:dGamma}
\dd^8\Gamma_i
& = & C_i\Bigl[X_i +\sin(2 \chi_{1}) \sin(2 \chi_{2})Y_i \Bigr]
\dLips(m^2;q_1,q_2,q_3,q_4) , \qquad i=H,A
\end{eqnarray}
where $H$ denotes a $CP$-even Higgs particle and $A$ a
$CP$-odd one.  Furthermore,
\begin{eqnarray}
\label{eq:invar}
X_{\rm H}
& = &
(q_1\cdot q_3)(q_2\cdot q_4) +(q_1\cdot q_4)(q_2\cdot q_3) \nn \\
Y_{\rm H}
& = &
(q_1\cdot q_3)(q_2\cdot q_4) -(q_1\cdot q_4)(q_2\cdot q_3) \nn \\
X_{\rm A}
& = &
-2[(q_1\cdot q_2)(q_3\cdot q_4)]^2 \nn \\
&   &
-2[(q_1\cdot q_3)(q_2\cdot q_4) -(q_1\cdot q_4)(q_2\cdot q_3)]^2 \nn \\
&   &
+(q_1\cdot q_2)(q_3\cdot q_4)
\{[(q_1\cdot q_3)+(q_2\cdot q_4)]^2
+[(q_1\cdot q_4)+(q_2\cdot q_3)]^2\} \nn \\
Y_{\rm A}
& = &
(q_1\cdot q_2)(q_3\cdot q_4)
[(q_1-q_2)\cdot(q_3+q_4)][(q_3-q_4)\cdot(q_1+q_2)]
\end{eqnarray}

These decay rates depend on the relative azimuthal orientation of the
planes formed by the decay products, in a way which reveals
the \cp quantum number of the decaying particle.
It is natural to define an azimuthal angle by
\begin{equation}
\cos \phi = \frac{\left({\bf {q}_{1}} \times {\bf {q}_{2}}\right)
            \cdot \left({\bf {q}_{3}} \times {\bf {q}_{4}}\right)}
                      {|{\bf {q}_{1}} \times {\bf {q}_{2}}|
                       |{\bf {q}_{3}} \times {\bf {q}_{4}}|}
\end{equation}
and consider the corresponding decay distributions.
(This is completely analogous to the classic determination of the
$\pi^0$ parity \cite{Yang50}.)
Integrating over the way the energy is shared within each pair,
over the invariant masses of the two pairs and the polar
angle distribution, we are left with a dependence on the azimuthal
angle $\phi$ only.
The term $Y_{\rm A}$ does not contribute to the resulting decay
distributions, which can be expressed in the following compact
way \citer{others,HagiwaraStong}
\begin{eqnarray}
\frac{2 \pi}{\Gamma_{\rm H}}\:\frac{\dd\Gamma_{\rm H}}{\dd\phi}
& = &
1 - \alpha(m)\sin(2 \chi_{1})\sin(2 \chi_{2}) \cos \phi
+ \beta(m)\cos 2 \phi \nn \\
\frac{2 \pi}{\Gamma_{\rm A}}\:\frac{\dd\Gamma_{\rm A}}{\dd\phi}
& = &
1 - \frac{1}{4} \cos 2 \phi
\end{eqnarray}
where the $\sin(2\chi)$ factors are given by ratios between the axial
and vector couplings \cite{skjosl93}.
The coefficients $\alpha(m)$ and $\beta(m)$ depend on whether
the decay takes place via $W^+W^-$ or via $ZZ$, and specifically
on the ratio of the vector boson mass to the Higgs mass $m$.
Representative distributions \cite{skjosl93} are shown in fig.~1
(left part).
Of course, jet identification is needed for this kind of analysis.
\begin{figure}[thb]
\begin{center}
\setlength{\unitlength}{1cm}
\begin{picture}(16,8.0)
\put(0.0,-1.5){
\mbox{\epsfysize=10cm\epsffile{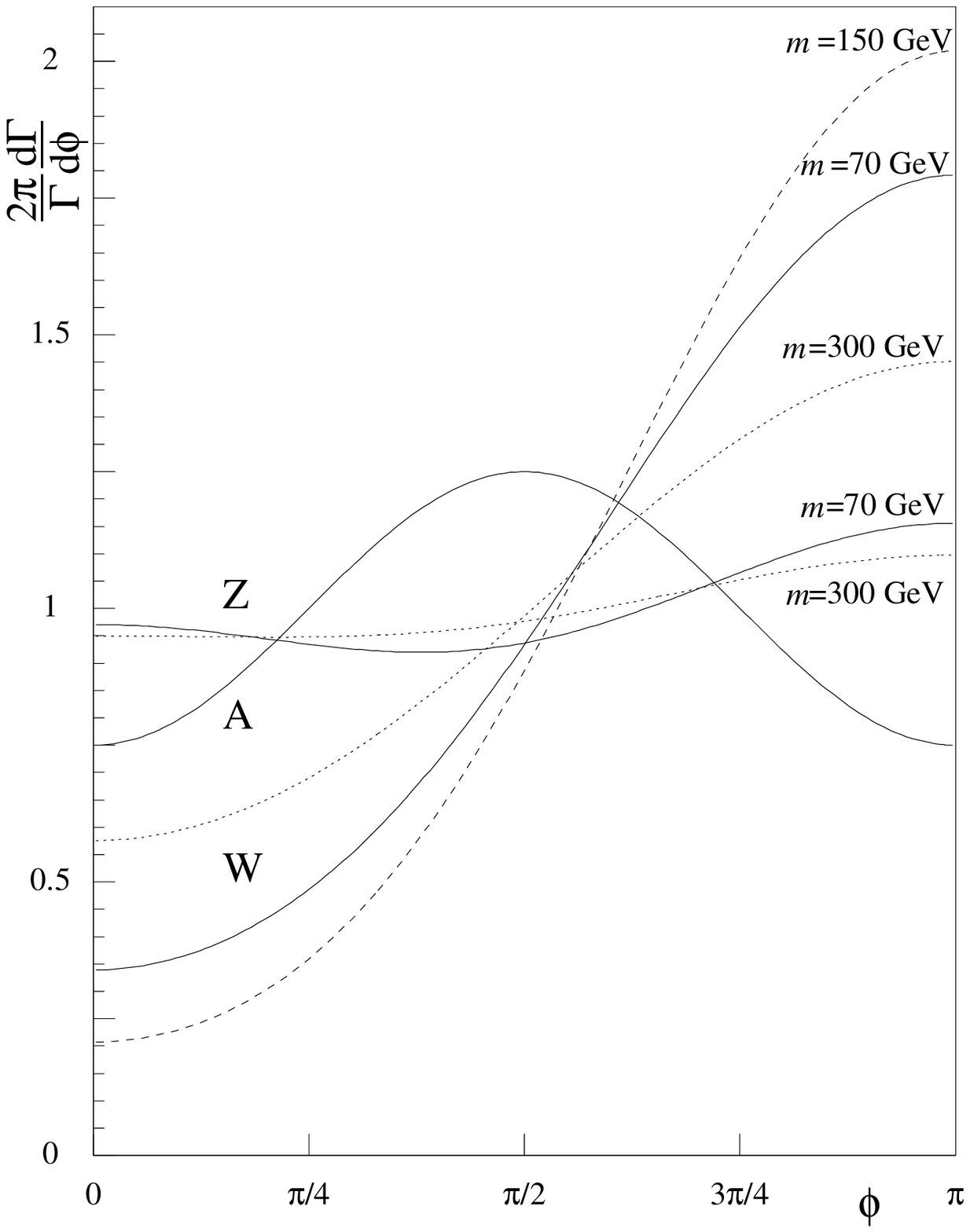}}
\mbox{\epsfysize=10cm\epsffile{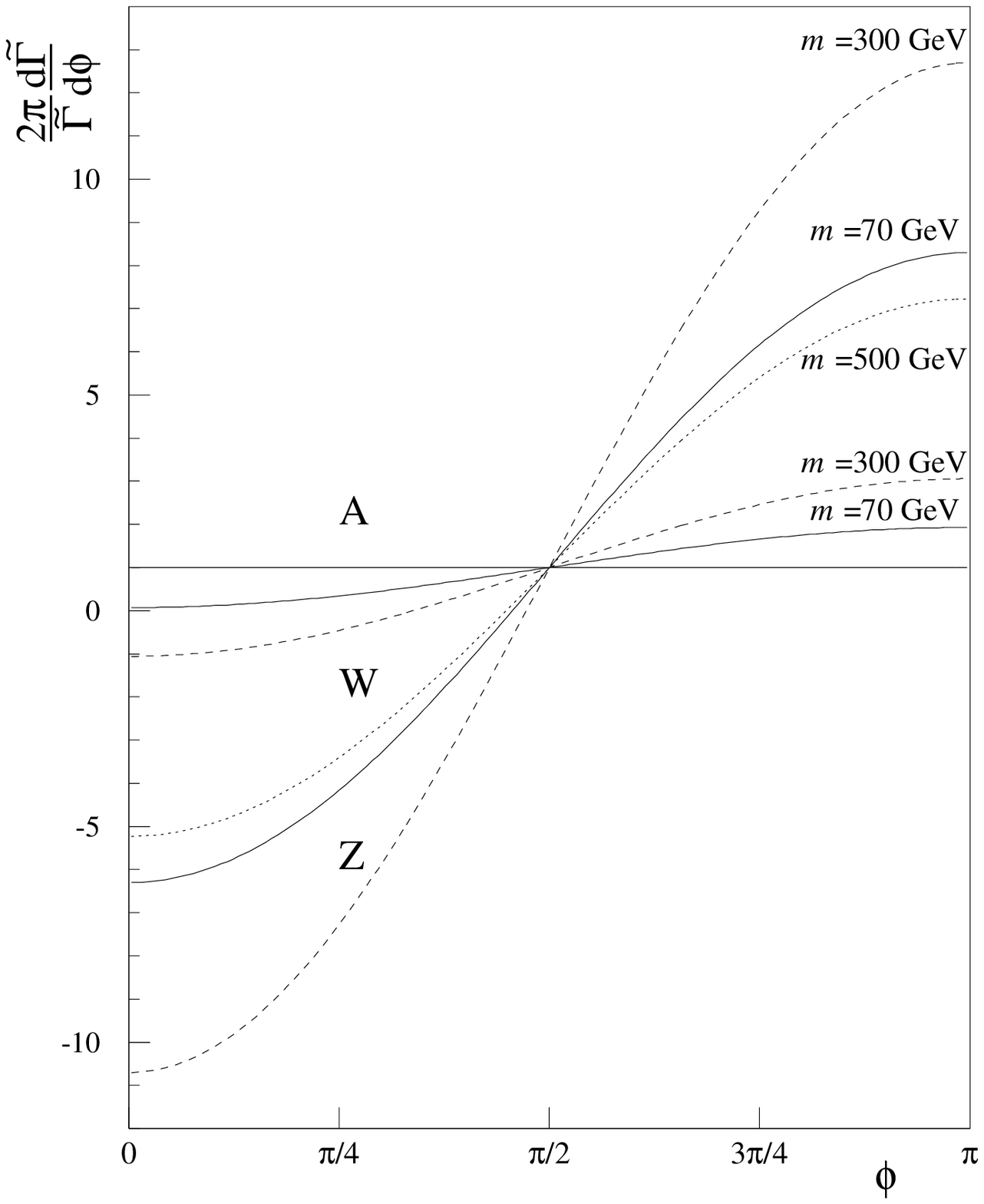}}}
\end{picture}
\begin{capt}
{\em Left part:}
Characteristic angular distributions of the planes defined by two
Dalitz pairs for CP-even Higgs particles decaying via two ${\rm W}$'s or two
${\rm Z}$'s, compared with the corresponding distribution for a CP-odd Higgs
particle (denoted ${\rm A}$). Different Higgs masses are considered in the
CP-even case.
{\em Right part:}
Characteristic energy-weighted distributions for CP-even Higgs
particles decaying via two ${\rm W}$'s or two
${\rm Z}$'s, compared with the corresponding distribution for a CP-odd Higgs
particle (${\rm A}$). (Ref.~\cite{skjosl93} had a sign error
in these distributions.)
\end{capt}
\end{center}
\end{figure}

In order to project out the $Y_{\rm A}$-term of eq.~(\ref{eq:invar}),
we multiply (\ref{eq:dGamma}) by the energy differences
$(\omega_{1}-\omega_{2})(\omega_{3}-\omega_{4})$ before
integrating over energies.
The distribution then takes the form \cite{skjosl93}
\begin{equation}
\frac{2 \pi}{\tilde{\Gamma}}
\:\frac{\dd\tilde{\Gamma}}{\dd\phi}
 =
1 - \frac{\kappa\left(m\right)}{\sin(2 \chi_{1})\sin(2 \chi_{2})}
\cos \phi
\end{equation}
in the $CP$-even case, whereas it is constant, $\kappa(m)=0$, in the
$CP$-odd case. Illustrative distributions are given in fig.~1
(right part).
This energy-weighted angular distribution provides an independent
determination of the \cp of the Higgs candidate, and is in fact more
sensitive than the unweighted angular distributions, especially
for heavy Higgs bosons.

\begin{figure}[thb]
\begin{center}
\setlength{\unitlength}{1cm}
\begin{picture}(16,7.0)
\put(4.0,-1.5)
{\mbox{\epsfysize=9.5cm\epsffile{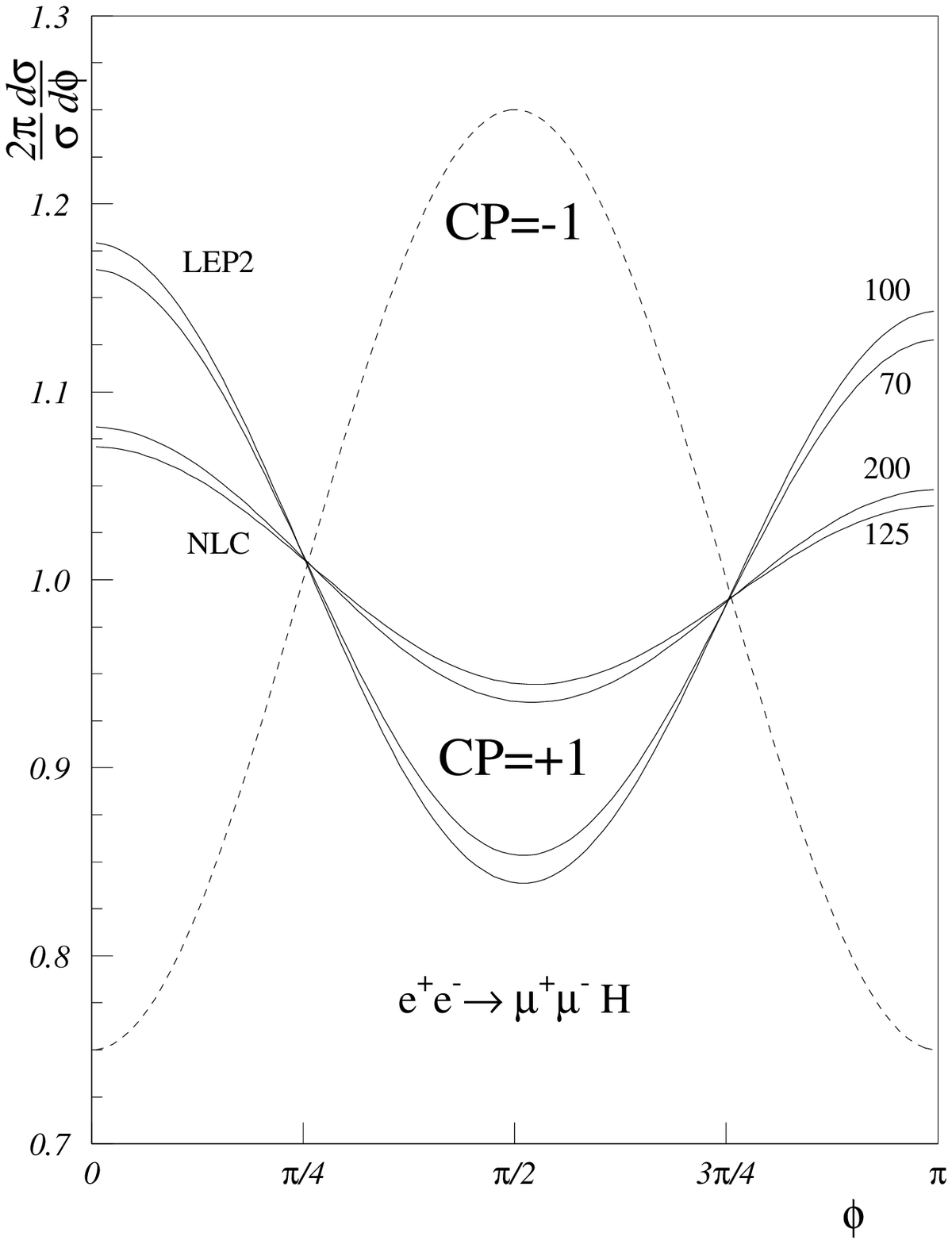}}}
\end{picture}
\begin{capt}
\label{bjlepnlc}
Angular distributions of the planes defined by
incoming $e^-$ and final-state fermi\-ons for a $CP$-even Higgs particle
(solid) compared with the corresponding distribution for a
$CP$-odd one (dashed).
Different energies and masses are considered in the $CP$-even case.
We assume $\sqrt{s}=200$ and 500~GeV at LEP2 and NLC, respectively.
\end{capt}
\end{center}
\end{figure}

Consider next the Bjorken process \cite{Bjorken}
\begin{equation}
e^{-}\left(p_{1}\right) e^{+}\left(p_{2}\right)
\rightarrow Z\left(Q\right) {h}\left(q_{3}\right)
\rightarrow f\left(q_{1}\right) {\bar f\left(q_{2}\right)}
{h}\left(q_{3}\right)
\label{EQU:Bj1}
\end{equation}
Let the momenta of the two final-state fermions and the initial electron
(in the overall {\it c.m.}\ frame)
define two planes,
and denote by $\phi$ the angle between those two planes; i.e.
\begin{equation}
\cos \phi = \frac{\left({\bf {p}_{1}} \times {\bf {Q}}\right)
            \cdot \left({\bf {q}_{1}} \times {\bf {q}_{2}}\right)}
                      {|{\bf {p}_{1}} \times {\bf {Q}}|
                       |{\bf {q}_{1}} \times {\bf {q}_{2}}|}
\label{EQU:Dj4}
\end{equation}
We shall discuss the angular distribution of the cross section
$\sigma$,
\begin{equation}
\frac{2\pi}{\sigma}\:
\frac{\dd\sigma}{\dd\phi}
\label{EQU:intro1}
\end{equation}
both in the case of $CP$-even and $CP$-odd Higgs bosons.
These distributions take the form
\cite{others,HagiwaraStong,skjosl95,Kniehl}
\begin{eqnarray}
\frac{2 \pi}{\sigma_H}\:\frac{\dd\sigma_H}{\dd\phi}
&=&
1 + \alpha(s,m) \cos \phi
  + \beta(s,m) \cos 2\phi,
\label{EQU:Dl50}
\\
\frac{2 \pi}{\sigma_A}\:\frac{\dd\sigma_A}{\dd\phi}
&=&
1 -\frac{1}{4} \cos 2\phi
\label{EQU:Dl51}
\end{eqnarray}
with typical cases shown in fig.~\ref{bjlepnlc}.
There is a clear difference between $CP=1$ and $CP=-1$ for
a wide range of energies and Higgs masses.

Experimentally, however, one faces the challenge of contrasting
two angular distributions with a restricted number of events
and allowing also for background.
We shall here focus on the intermediate Higgs mass range;
more specifically, we consider $m\lsim 140$~GeV where the Higgs
decays dominantly to $b\bar{b}$.
The main background will then stem from
$e^+e^- \rightarrow ZZ$ and also
$e^+e^- \rightarrow Z\gamma, \gamma \gamma$.
The cleanest channel for isolating the Higgs signal from the background
is provided by the $\mu^+\mu^-$ and
$e^+ e^-$ decay modes of the $Z$ boson.

Let us next limit consideration to the energy range
$\sqrt{s}=300-500$~GeV, as appropriate for a linear collider
\cite{Wiik}, henceforth denoted NLC.
We impose reasonable cuts and constraints as described in
\cite{Kniehl}; e.g. $|m_{\mu^+\mu^-}-m_Z| \leq 6$~GeV and
$|\cos\theta_Z| \leq 0.6$.
The signal for
$e^+e^- \rightarrow Z H \rightarrow \mu^+\mu^- b \bar{b}$ will then
be larger than the background
$e^+e^- \rightarrow Z Z \rightarrow \mu^+\mu^- b \bar{b}$
by an order of magnitude.
In the following we shall thus neglect the background
in the discussion of (\ref{EQU:Dl50}) versus (\ref{EQU:Dl51}).
With  $\sigma(e^+e^- \rightarrow Z H) \sim 200$~fb and an integrated
luminosity of 20 ${\rm fb}^{-1}$ a year \cite{Kniehl},
about 4000 Higgs particles will be produced per year, in this
intermediate mass range.
However, following~\cite{Kniehl} we have only $\sim 30$ signal events
$e^+e^- \rightarrow Z H \rightarrow \mu^+\mu^- b \bar{b}$ left
per year for e.g.\ a NLC operating at $\sqrt{s}=300$~GeV
and a Higgs particle of mass $m=125$~GeV.
In the case $e^+e^- \rightarrow Z H \rightarrow e^+ e^- b \bar{b}$
we also have a t-channel background contribution
from the $ZZ$ fusion process
$e^+e^- \rightarrow e^+e^- (Z Z) \rightarrow e^+ e^- H$.
This contribution may be neglected at LEP energies,
but it is comparable to the s-channel contribution at higher
energies. However, this contribution can be suppressed
by imposing a cut on the invariant mass of the final-state electrons,
e.g.\ $|m_{e^+e^-}-m_Z| \leq 6$~GeV.
Hence, we can effectively treat the electrons on the same footing as
the muons, thereby obtaining a doubling of the event rate.

In order to demonstrate the potential of the NLC for determining
the $CP$ of the Higgs particle,
we show in fig.~3 the result of a Monte Carlo simulation.
For this purpose we have used PYTHIA \cite{Sjostrand},
suitably modified to allow for the $CP$-odd case.
The statistics correspond to 3~years of running\footnote{The event
rate is based on the Standard Model, and could be different for
a non-standard Higgs sector.}
using both the $\mu^+\mu^-$ and $e^+ e^-$ decay modes of the $Z$ boson.
This yields about 200 events in these channels,
assuming an integrated luminosity of 20~fb${}^{-1}$ per year.

\begin{figure}[thb]
\begin{center}
\setlength{\unitlength}{1cm}
\begin{picture}(16,6.5)
\put(1.0,-1.5){\mbox{\epsfysize=9cm\epsffile{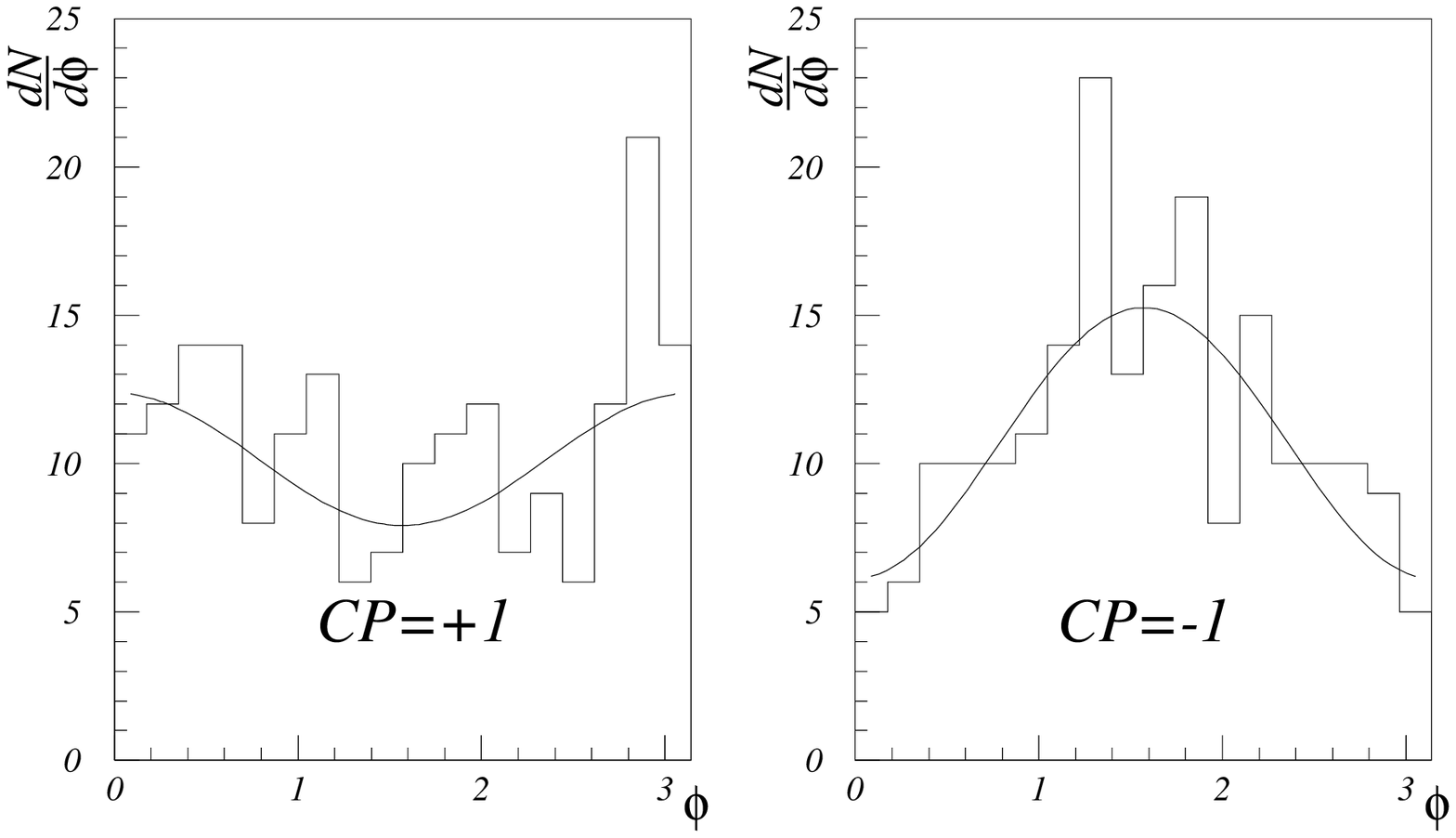}}}
\end{picture}
\begin{capt}
Monte Carlo data
displaying the angular distribution of events
$e^+e^- \rightarrow Z H \rightarrow l^+l^- b \bar{b}$, $l=\mu,e$
for a Standard-Model Higgs versus a CP-odd one.
We have taken $\sqrt{s}=300$~GeV, $m=125$~GeV, and an angular cut
$|\cos\theta|\le0.6$.
\end{capt}
\end{center}
\end{figure}
For $\sqrt{s}=300~\GeV$ and $m_H=125~\GeV$,
the `bare' prediction for $\beta$ is 0.12.
The cut on polar angle, $|\cos\theta|\le0.6$,
increases it slightly to 0.14.
Similarly, the `$-1/4$' changes to $-0.39$.
Thus, the cut makes it easier
to discriminate between the $CP$-even distribution
and the $CP$-odd one \cite{skjosl95}.
{}From fig.~3 we see that the individual angular
Monte Carlo distributions are consistent with the predictions,
showing that a three-year data sample is large enough to
reproduce the azimuthal distributions.
In the Standard-Model case the fit gives $0.92\pm 0.07$ and
$0.2\pm 0.1$ for the predictions 1.00 and 0.14, respectively,
with $\chi^2=1.0$.
In the $CP$-odd case the fit gives $0.94\pm 0.07$ and
$-0.4\pm 0.1$ for the predictions 1.00 and $-0.39$, respectively,
with $\chi^2=0.7$.
More importantly, since the $\cos 2\phi$ terms are more than 4 standard
deviations away from each other, a data sample of this size is sufficient
to verify the scalar nature of the Standard-Model Higgs.
Using likelihood ratios, as described in \cite{Roe}, for choosing between the
two hypotheses of $CP$ even and $CP$ odd, we find that less than 3 years of
running suffices.

An alternative test has recently been suggested
by Arens et.~al.~\cite{Arens} in the context of Higgs
decaying via vector bosons to four fermions, where one studies
the energy spectrum of one of the final-state fermions.
Applying this idea to the Bjorken process one would
study the energy distribution of an outgoing fermion, e.g.\ $\mu^-$ or $e^-$.
Introducing the scaled lepton energy, $x=4E_{l^-}/\sqrt{s}$, $l=\mu,e$,
we shall consider the energy distribution of the cross section
with respect to this final-fermion energy,
\begin{equation}
\frac{1}{\sigma}\:
\frac{\dd\sigma}{\dd x}
\label{EQU:intro2}
\end{equation}
both in the case of $CP$-even and $CP$-odd Higgs bosons. We are using
the narrow-width approximation and the range of $x$ is given by
$x_-\le x\le x_+$, with $sx_{\pm}=s+m_Z^2-m^2\pm\sqrt{\lambda}$.
Here the distributions are given as second-degree polynominals
in $x$, and, as shown in~\cite{skjosl95}, the coefficients have a non-trivial
dependence on the {\it c.m.}\  energy and the Higgs mass, also for the
$CP$-odd case.
A representative set of energy distributions is given in
fig.~4 for the case $e^+e^- \rightarrow \mu^+\mu^- h$
for both LEP2 and NLC energies.
Again, there is a clear difference between the $CP$-even
and the $CP$-odd cases.

A Monte-Carlo simulation for the energy distribution eq.~(\ref{EQU:intro2})
confirms this statement \cite{skjosl95}.
An analysis of likelihood ratios demonstrates that less than
3 years of running is sufficient if we require
a discrimination by four standard deviations,
but more events seem to be required than in the case of angular
distributions.

\begin{figure}[htb]
\begin{center}
\setlength{\unitlength}{1cm}
\begin{picture}(16,6.0)
\put(1.,-1.5){\mbox{\epsfysize=9cm\epsffile{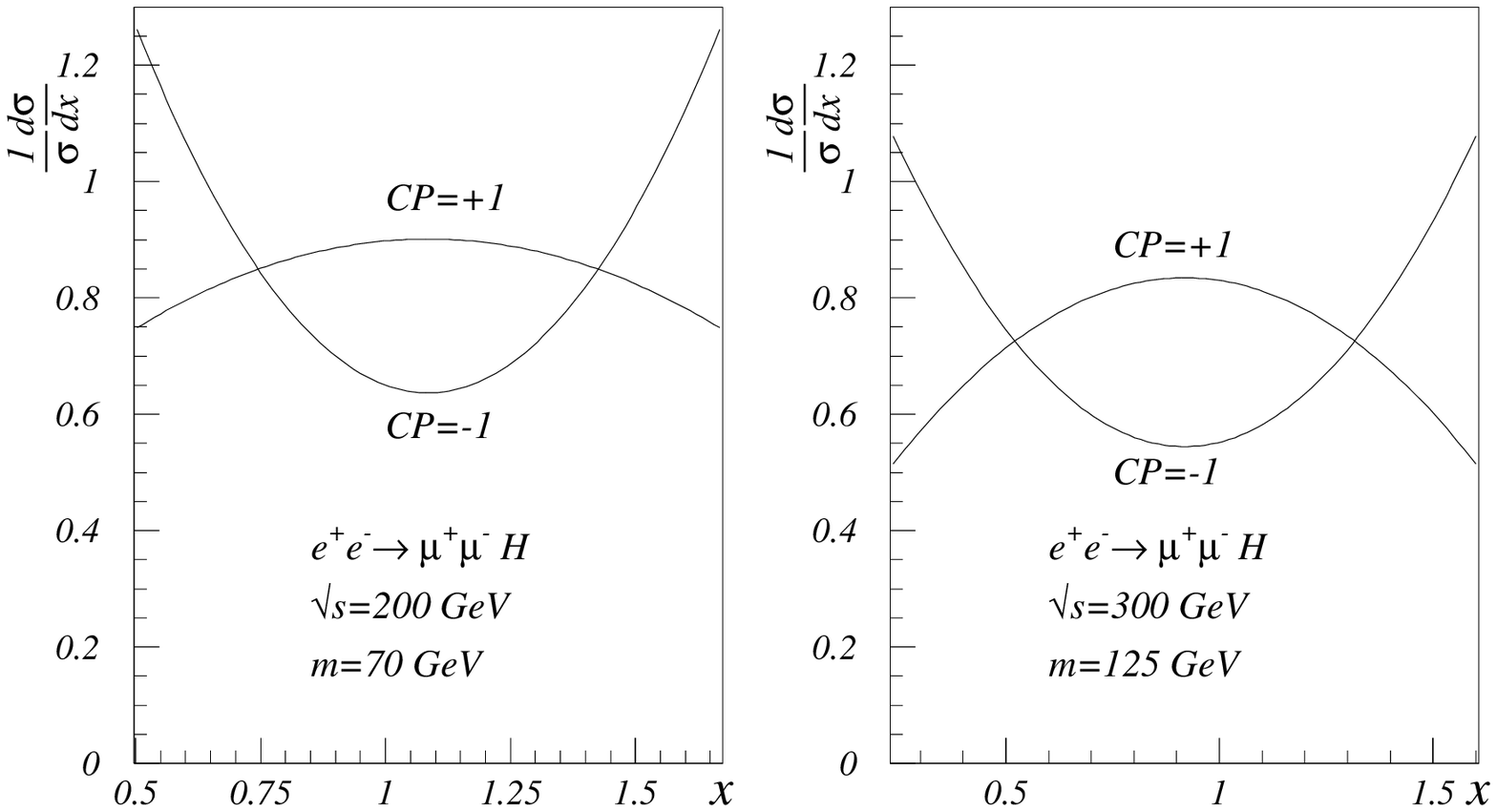}}}
\end{picture}
\begin{capt}
\label{bjspec}
Characteristic distributions
for the scaled energy, eq.~(\ref{EQU:intro2}), of the $l^-$,
$l=\mu,e$ in the Bjorken process $e^+e^- \rightarrow l^+l^- h$.
Different energies and masses are considered.
\end{capt}
\end{center}
\end{figure}
\section{$CP$ Violation}
If we allow for both the Standard-Model and the $CP$-odd term
in the Higgs-vector coupling (\ref{eq:vertex}),
then there will be $CP$ violation in Higgs decay of the kind
(\ref{eq:LHC}) as well as in the Bjorken process
\cite{HagiwaraStong,skjosl95,Kniehl,othersCP,skjosl94}.
The decay rate can be written as
\begin{equation}
\dd^8\Gamma
= \sqrt{2}\, \frac{G_{\rm F}}{m}\,
N_1 N_2 D\biggl[X + \frac{\eta}{m_V^2}\, Y
+ \left(\frac{\eta}{m_V^2}\right)^{2} \, Z \biggr]
\dLips(m^2;q_1,q_2,q_3,q_4)
\end{equation}
where the $Y$ term describes \cp violation.

Integrating over the way in which energy is shared in each
fermion-antifermion pair,
as well as over polar angles,
we can write
the distribution in a compact form as \cite{skjosl94}
\begin{eqnarray}
 \frac{\dd^3\Gamma }{\dd\phi\: \dd s_1 \dd s_2}
& = &\frac{\sqrt{2}}{72 (4\pi)^6}\, N_1 N_2 \, \frac{G_{\rm F}}{m^3}
\sqrt{\lambda\left(m^2,s_1,s_2\right)}\, D(s_1,s_2)\  \nn \\
& & \times
\biggl[ \lambda\left(m^2,s_1,s_2\right)+4 s_1 s_2 \left(1+2 \rho^{2} \right)
+2s_1 s_2\,\rho^2\,\cos 2(\phi - \delta) \nn \\
& & -  \sin 2 \chi_{1} \sin 2 \chi_{2} \left(\frac{3 \pi}{4}\right)^2
\sqrt{s_1 s_2}\, (m^2-s_1-s_2)\,\rho\,\cos (\phi - \delta) \biggr]
\end{eqnarray}
with a modulation function $\rho$
\begin{equation}
\rho=\sqrt{1+\eta^2\lambda\left(m^2,s_1,s_2\right)/(4m_V^4)}
\end{equation}
and a ``tilt" given by the amount of \cp violation as
\begin{equation}
\delta=
\arctan\frac{\eta(s_1,s_2)\sqrt{\lambda(m^2,s_1,s_2)}}{2m_V^2},
\qquad
-\pi/2 < \delta < \pi/2
\end{equation}
and $\lambda(x,y,z)=x^2+y^2+z^2-2(xy+xz+yz)$ the K{\"{a}}llen function.

An inclusive distribution is obtained if we integrate
also over the invariant masses of the two pairs.
It has the form
\begin{equation}
\frac{2 \pi}{\Gamma}\:\frac{\dd\Gamma}{\dd\phi}
= 1 - \alpha(m) \, \rho \, \cos (\phi -\delta)
+ \beta(m) \, \rho^{2} \, \cos 2\left(\phi -\delta\right)
\end{equation}

The tilt $\delta$ will depend on the mass of the Higgs particle,
and is shown for some representative values of $\eta$
in fig.~5 (left part).
If the Higgs is somewhat heavy, and $\eta$ not too small,
this tilt can be significant.
Again, energy-weighted correlations provide an independent
way of measuring this kind of \cp violation
\cite{skjosl94}.

The same kind of analysis may be applied to the Bjorken process.
There is a relative shift in the angular distribution
of the two planes, defined by eq.~(\ref{EQU:Dj4}),
due to $CP$ violation.
This rotation vanishes at the threshold for producing
a real vector boson and,
even for a fixed value of $\eta$, grows with energy.
The distribution takes the compact form
\begin{equation}
\label{EQU:Dl5}
\frac{2 \pi}{\sigma}\:\frac{\dd\sigma}{\dd\phi}
= 1 + \alpha(s,m) \, \rho \, \cos (\phi +\delta)
+ \beta(s,m) \, \rho^{2} \, \cos 2\left(\phi +\delta\right)
\end{equation}
Any $CP$ violation would thus show up as a tilt in the
azimuthal distribution, by the amount $\delta$.

\begin{figure}[thb]
\begin{center}
\setlength{\unitlength}{1cm}
\begin{picture}(16,8.0)
\put(0.0,-1){\mbox{\epsfysize=10cm\epsffile{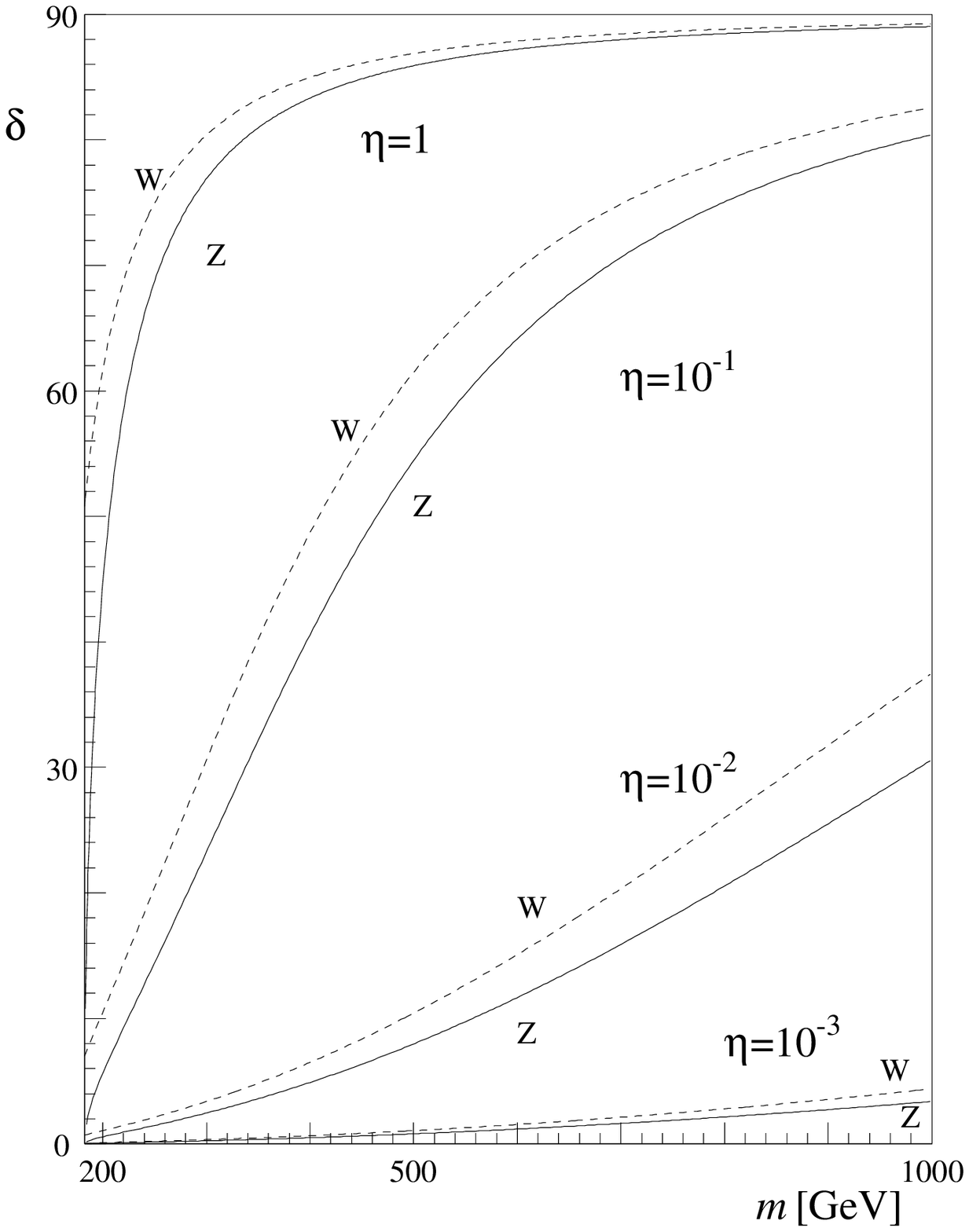}}
             \mbox{\epsfysize=10cm\epsffile{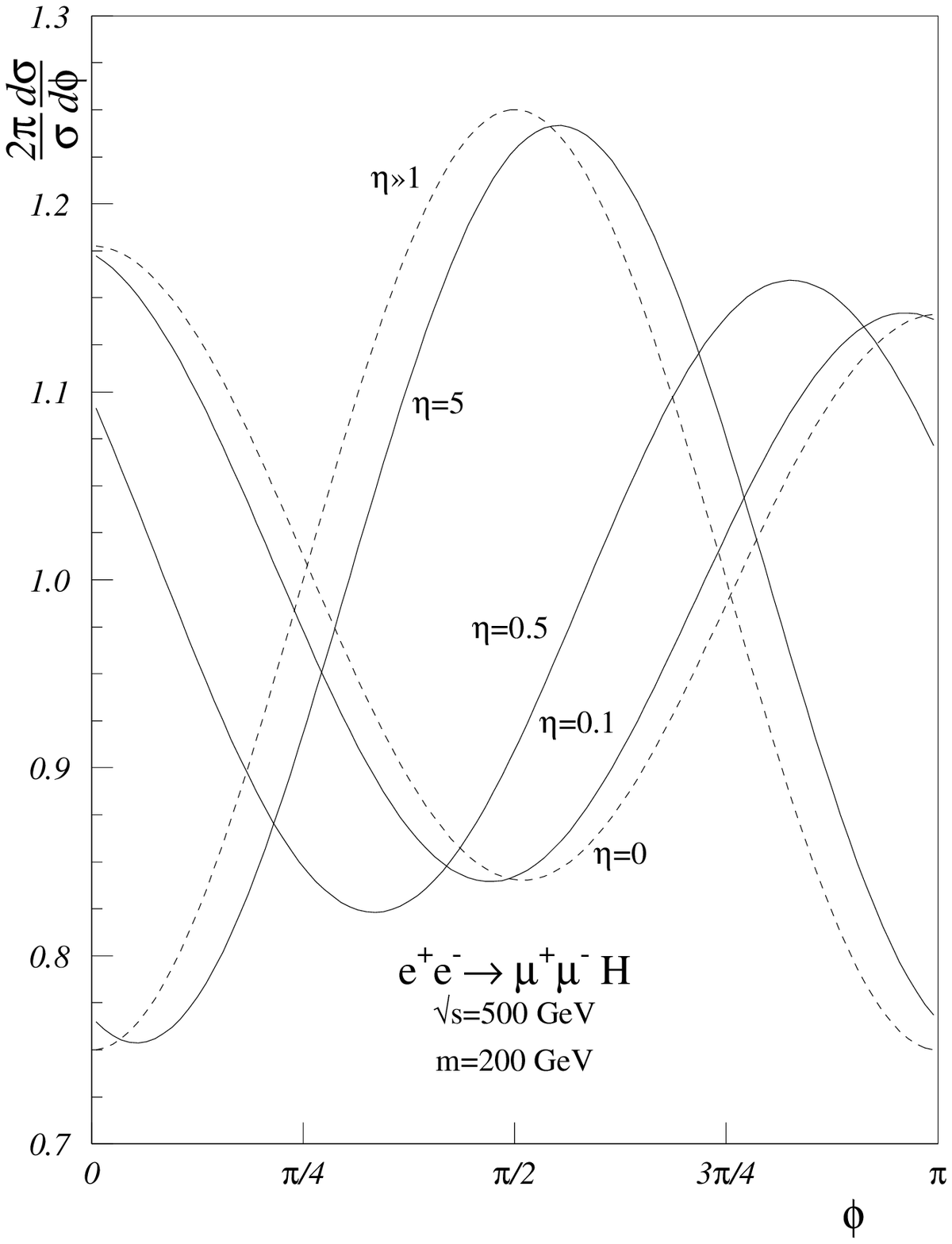}}}
\end{picture}
\begin{capt}
{\em Left part:}
The angle $\delta$ (in degrees) for a Higgs particle
of mass $m$, for $\eta=1$, $10^{-1}$, $10^{-2}$, and $10^{-3}$.
{\em Right part:}
Characteristic angular distributions
for different amounts of $CP$ violation in the Bjorken process,
including the $CP$-even ($\eta=0$) and $CP$-odd ($\eta\gg1$)
eigenstates. We have used
$\eta=0.1, 0.5, 5$ for $\sqrt{s}=500$~GeV and $m=200$~GeV.
\end{capt}
\end{center}
\end{figure}
A representative set of angular distributions is given in
fig.~5 (right part) for a broad range of $\eta$ values.
We have considered a Higgs boson of $m=200$~GeV
accompanied by a $\mu^{+} \mu^{-}$-pair in the final state,
produced at $\sqrt{s}=500$~GeV.
We observe that for $\eta \lsim 0.1$ and $\eta \gsim 5$,
the deviations from the $CP$-even and $CP$-odd distributions,
respectively, are small.
Experimentally it will be very difficult to disentangle two
distributions which differ by such a small phase shift.
Thus, observation of a small amount
of $CP$ violation would require a very large amount of data.

We note that the special cases $\eta=0$ and $\eta \gg 1$
correspond to the $CP$ even and $CP$ odd eigenstates,
respectively. Hence, the distribution (\ref{EQU:Dl5})
should be interpreted as being intermediate between those for
the two eigenstates.

\section{Summary}
We have discussed ways to analyze the \cp properties
of Higgs candidates through angular decay distributions.
In particular,
we have addressed the problem of estimating the amount of data needed
in order to distinguish a scalar Higgs from a pseudoscalar one
at a future linear collider.
This is most likely not possible at LEP2 due to much smaller event rates and
background which is not easily suppressed.
However, we have demonstrated that one will be able to establish
the scalar nature of the Higgs boson at the Next Linear Collider
from an analysis of angular or energy correlations.
This particular study has been carried out for the case $\sqrt{s}=300$~GeV,
$m=125$~GeV. Similar results are expected in other cases as long as
the background is small.

In order to establish or rule out specific models, one will also need
to compare different branching ratios, in particular to fermionic
final states.  The methods discussed above instead deal with quite
general properties of the models.

\medskip
We are grateful to the Organizers of the Third
Tallinn Symposium on Neutrino Physics,
in particular Professor I. Ots,
for creating a stimulating and pleasant atmosphere during the meeting.
This research has been supported by the Research Council of Norway.

\end{document}